\documentclass{aastex63}

\usepackage[utf8x]{inputenc}
\usepackage{amssymb}
\usepackage{amsmath}
\usepackage{natbib}
\usepackage{hyperref}

\received{\today}
\revised{\today}
\accepted{\today}
\submitjournal{ApJ}

\shorttitle{Does Turbulence along the CS drive ICWs?}
\shortauthors{Telloni et al.}

\begin{document}

\title{Does Turbulence along the Coronal Current Sheet Drive Ion Cyclotron Waves?}

\correspondingauthor{Daniele Telloni}
\email{daniele.telloni@inaf.it}

\author[0000-0002-6710-8142]{Daniele Telloni}
\affil{National Institute for Astrophysics, Astrophysical Observatory of Torino, Via Osservatorio 20, I-10025 Pino Torinese, Italy}
\author[0000-0002-4642-6192]{Gary P. Zank}
\affil{Center for Space Plasma and Aeronomic Research, University of Alabama in Huntsville, Huntsville, AL 35805, USA}
\affil{Department of Space Science, University of Alabama in Huntsville, Huntsville, AL 35805, USA}
\author[0000-0003-1549-5256]{Laxman Adhikari}
\affil{Center for Space Plasma and Aeronomic Research, University of Alabama in Huntsville, Huntsville, AL 35805, USA}
\author[0000-0002-4299-0490]{Lingling Zhao}
\affil{Center for Space Plasma and Aeronomic Research, University of Alabama in Huntsville, Huntsville, AL 35805, USA}
\author[0000-0002-1017-7163]{Roberto Susino}
\affil{National Institute for Astrophysics, Astrophysical Observatory of Torino, Via Osservatorio 20, I-10025 Pino Torinese, Italy}
\author[0000-0003-4155-6542]{Ester Antonucci}
\affil{National Institute for Astrophysics, Astrophysical Observatory of Torino, Via Osservatorio 20, I-10025 Pino Torinese, Italy}
\author[0000-0002-2789-816X]{Silvano Fineschi}
\affil{National Institute for Astrophysics, Astrophysical Observatory of Torino, Via Osservatorio 20, I-10025 Pino Torinese, Italy}
\author[0000-0002-5365-7546]{Marco Stangalini}
\affil{Italian Space Agency, Via del Politecnico snc, I-00133 Roma, Italy}
\author[0000-0002-5467-6386]{Catia Grimani}
\affil{University of Urbino Carlo Bo, Department of Pure and Applied Sciences, Via Santa Chiara 27, I-61029 Urbino, Italy}
\affil{National Institute for Nuclear Physics, Section in Florence, Via Bruno Rossi 1, I-50019 Sesto Fiorentino, Italy}
\author[0000-0002-5981-7758]{Luca Sorriso-Valvo}
\affil{National Research Council, Institute for the Science and Technology of Plasmas, Via Amendola 122/D, I-70126 Bari, Italy}
\affil{Swedish Institute of Space Physics, \AA ngstr\"om Laboratory, L\"agerhyddsv\"agen 1, SE-751 21 Uppsala, Sweden}
\author[0000-0002-0497-1096]{Daniel Verscharen}
\affil{Mullard Space Science Laboratory, University College London, Holmbury St. Mary, RH5 6NT Dorking, UK}
\author[0000-0002-6433-7767]{Raffaele Marino}
\affil{Universit\'e de Lyon, Centre National de la Recherche Scientifique, \'Ecole Centrale de Lyon, Institut National des Sciences Appliqu\'ees de Lyon, Universit\'e Claude Bernard Lyon 1, Laboratoire de M\'ecanique des Fluides et d'Acoustique, F-69134 \'Ecully, France}
\author[0000-0002-3468-8566]{Silvio Giordano}
\affil{National Institute for Astrophysics, Astrophysical Observatory of Torino, Via Osservatorio 20, I-10025 Pino Torinese, Italy}
\author[0000-0003-2647-117X]{Raffaella D'Amicis}
\affil{National Institute for Astrophysics, Institute for Space Astrophysics and Planetology, Via del Fosso del Cavaliere 100, I-00133 Roma, Italy}
\author[0000-0003-1059-4853]{Denise Perrone}
\affil{Italian Space Agency, Via del Politecnico snc, I-00133 Roma, Italy}
\author[0000-0002-3559-5273]{Francesco Carbone}
\affil{National Research Council, Institute of Atmospheric Pollution Research, c/o University of Calabria, I-87036 Rende, Italy}
\author[0000-0002-0016-7594]{Alessandro Liberatore}
\affil{Jet Propulsion Laboratory, California Institute of Technology, Pasadena, CA 91109, USA}
\author[0000-0002-2152-0115]{Roberto Bruno}
\affil{National Institute for Astrophysics, Institute for Space Astrophysics and Planetology, Via del Fosso del Cavaliere 100, I-00133 Roma, Italy}
\author[0000-0002-9207-2647]{Gaetano Zimbardo}
\affil{University of Calabria, Department of Physics, Ponte Pietro Bucci Cubo 31C, I-87036 Rende, Italy}
\author[0000-0001-9921-1198]{Marco Romoli}
\affil{University of Florence, Department of Physics and Astronomy, Via Giovanni Sansone 1, I-50019 Sesto Fiorentino, Italy}
\author[0000-0003-1962-9741]{Vincenzo Andretta}
\affil{National Institute for Astrophysics, Astronomical Observatory of Capodimonte, Salita Moiariello 16, I-80131 Napoli, Italy}
\author[0000-0001-6273-8738]{Vania Da Deppo}
\affil{National Research Council, Institute for Photonics and Nanotechnologies, Via Trasea 7, I-35131 Padova, Italy}
\author[0000-0002-5778-2600]{Petr Heinzel}
\affil{Czech Academy of Sciences, Astronomical Institute, Fri\v{c}ova 298, CZ-25165 Ond\v{r}ejov, Czechia}
\author[0000-0001-9670-2063]{John D. Moses}
\affil{National Aeronautics and Space Administration, Headquarters, Washington, DC 20546, USA}
\author[0000-0003-2007-3138]{Giampiero Naletto}
\affil{University of Padua, Department of Physics and Astronomy, Via Francesco Marzolo 8, I-35131 Padova, Italy}
\author[0000-0002-9459-3841]{Gianalfredo Nicolini}
\affil{National Institute for Astrophysics, Astrophysical Observatory of Torino, Via Osservatorio 20, I-10025 Pino Torinese, Italy}
\author[0000-0003-3517-8688]{Daniele Spadaro}
\affil{National Institute for Astrophysics, Astrophysical Observatory of Catania, Via Santa Sofia 78, I-95123 Catania, Italy}
\author[0000-0001-7298-2320]{Luca Teriaca}
\affil{Max Planck Institute for Solar System Research, Justus-von-Liebig-Weg 3, D-37077 G\"ottingen, Germany}
\author[0000-0002-8734-808X]{Aleksandr Burtovoi}
\affil{National Institute for Astrophysics, Astrophysical Observatory of Arcetri, Largo Enrico Fermi 5, I-50125 Firenze, Italy}
\author[0000-0003-2426-2112]{Yara De Leo}
\affil{Max Planck Institute for Solar System Research, Justus-von-Liebig-Weg 3, D-37077 G\"ottingen, Germany}
\affil{University of Catania, Department of Physics and Astronomy, Via Santa Sofia 64, I-95123 Catania, Italy}
\author[0000-0002-0764-7929]{Giovanna Jerse}
\affil{National Institute for Astrophysics, Astronomical Observatory of Trieste, Localit\`a Basovizza 302, I-34149 Trieste, Italy}
\author[0000-0001-8244-9749]{Federico Landini}
\affil{National Institute for Astrophysics, Astrophysical Observatory of Torino, Via Osservatorio 20, I-10025 Pino Torinese, Italy}
\author[0000-0002-3789-2482]{Maurizio Pancrazzi}
\affil{National Institute for Astrophysics, Astrophysical Observatory of Torino, Via Osservatorio 20, I-10025 Pino Torinese, Italy}
\author[0000-0002-5163-5837]{Clementina Sasso}
\affil{National Institute for Astrophysics, Astronomical Observatory of Capodimonte, Salita Moiariello 16, I-80131 Napoli, Italy}
\author[0000-0001-7762-280X]{Alessandra Slemer}
\affil{National Research Council, Institute for Photonics and Nanotechnologies, Via Trasea 7, I-35131 Padova, Italy}

\begin{abstract}
Evidence for the presence of ion cyclotron waves, driven by turbulence, at the boundaries of the current sheet is reported in this paper. By exploiting the full potential of the joint observations performed by Parker Solar Probe and the Metis coronagraph on board Solar Orbiter, local measurements of the solar wind can be linked with the large-scale structures of the solar corona. The results suggest that the dynamics of the current sheet layers generates turbulence, which in turn creates a sufficiently strong temperature anisotropy to make the solar-wind plasma unstable to anisotropy-driven instabilities such as the Alfv\'en ion-cyclotron, mirror-mode, and firehose instabilities. The study of the polarization state of high-frequency magnetic fluctuations reveals that ion cyclotron waves are indeed present along the current sheet, thus linking the magnetic topology of the remotely imaged coronal source regions with the wave bursts observed in situ. The present results may allow improvement of state-of-the-art models based on the ion cyclotron mechanism, providing new insights into the processes involved in coronal heating.
\end{abstract}

\keywords{instabilities --- magnetohydrodynamics (MHD) --- turbulence --- waves --- Sun: corona --- solar wind}

\section{Introduction}
\label{sec:introduction}
The physical processes that heat the solar corona plasma to temperatures exceeding one million degrees, thereby driving the solar wind \citep{1972cesw.book.....H}, are still strongly debated. Solving this puzzle of the solar physics is central to the Parker Solar Probe \citep[PSP;][]{2016SSRv..204....7F} and Solar Orbiter \citep[SO;][]{2020A&A...642A...1M} missions. The many models proposed so far to explain the so-called ``coronal heating problem'' can be classified into two major categories, namely alternating current (AC) and direct current (DC) heating mechanisms. AC/DC models involve energy release by waves/turbulence or magnetic reconnection events \citep[as nanoflares,][]{1972ApJ...174..499P} respectively. Reviews can be found in \citet{1993SoPh..148...43Z} and \citet{2000ApJ...530..999M}. On the AC (wave/turbulence) side, two key mechanisms candidates are Ion Cyclotron Waves (ICWs) dissipation \citep{1999JGR...104.2521L,2000ApJ...532.1197C,2000ApJ...537.1054L,2002JGRA..107.1147H} and low-frequency MagnetoHydroDynamic (MHD) turbulence dissipation. The latter can further take place either through nonlinear coupling of a dominant outward Alfv\'en wave flux with a minority of inwardly reflected modes \citep{1999ApJ...523L..93M,2009ApJ...700L..39V,2010ApJ...708L.116V}, or, in the Nearly Incompressible (NI) MHD theory, in the frame of quasi-$2$D turbulence dynamics generated by the magnetic carpet and advected into the solar corona where it can then dissipate \citep{2017ApJ...835..147Z,2018ApJ...854...32Z,2020ApJ...900..115Z}. Although recent observational indications seem to point toward the interpretation in terms of the quasi-$2$D scenario \citep[the reader is referred to][where the above turbulence models are reviewed and their predictions tested against PSP observations]{2021PhPl...28h0501Z}, there is still no general consensus on which physical processes drive the deposition of the energy at small scales, needed to heat the plasma and accelerate the solar wind. Common to the aforementioned AC classifications, and indeed integral to energy transport and solar corona heating, are the collisionless field-particle interactions occurring at scales close to the proton inertial length and proton gyro-radius. At these scales, the solar wind, which has a fluid-like behavior in the inertial range of turbulence \citep[see the exhaustive review by][and references therein]{2013LRSP...10....2B}, takes on dispersive kinetic characteristics \citep{2005PhRvL..94u5002B,2008PhRvL.100f5004H,2010PhRvL.105m1101S,2013SSRv..178..101A}. Energy is thus transferred from the fluctuating fields to thermal particle energy, that is, dissipated, resulting in plasma heating.

PSP measurements compare remarkably well with the quasi-$2$D turbulence model developed by \citet{2017ApJ...835..147Z}, both in slow \citep{2020ApJS..246...38A,2022ApJ...938L...8T} and fast \citep{2020ApJ...901..102A,2021A&A...650A..16A} solar wind flows. In addition, the theoretical predictions based on the NI MHD theory do satisfactorily match the observed acceleration of coronal outflows to supersonic \citep{2007A&A...472..299T,2007A&A...476.1341T} and super-Alfv\'enic \citep{2022ApJ...937L..29A} speeds. Succeeding in reproducing observations of both coronal and heliospheric plasmas, the NI quasi-$2$D model is thus gaining reputation as a credible mechanism for solar wind plasma heating and acceleration. On the other hand, a number of fundamental major issues concerning the ion cyclotron heating mechanism \citep[see][for more details]{2021PhPl...28h0501Z} seem to rule it out from playing a relevant role in the interplanetary space plasma heating processes. A decade of remote-sensing observations of the extended corona with the UltraViolet Coronagraph Spectrometer \citep[UVCS;][]{1995SoPh..162..313K} on board the SOlar and Heliospheric Observatory \citep[SOHO;][]{1995SoPh..162....1D}, however, appear to agree well with a scenario in which ions/atoms are heated by cyclotron-resonant interaction with high-frequency Alfv\'en waves \citep{1998ApJ...501L.127K,1999ApJ...511..481C}. In particular, the kinetic temperatures of different coronal ion species were found to vary with their charge-to-mass $Z/A$ ratio \citep{1999ApJ...518..937C}, as expected for ion cyclotron wave dissipation, whose rate depends on the local ion gyro-frequency $\Omega_{i}\propto Z/A$. Similarly, in-situ measurements also strongly indicate that ICW damping is involved in the energy cascade \citep{1998ApJ...507L.181L,2014ApJ...787L..24B}. Evidence for the presence of ICWs in space plasmas \citep{2011ApJ...731...85H,2011ApJ...734...15P,2015ApJ...805...46T,2015ApJ...811L..17B,2022ApJ...928...36L} when the temperature anisotropy thresholds of the cyclotron instability are exceeded \citep{2009PhRvL.103u1101B,2016MNRAS.463L..79T,2019ApJ...884L..53W,2019ApJ...885L...5T} points to ion cyclotron resonance as a viable energy-conversion mechanism in collisionless plasmas \citep{2022PhRvL.129p5101B}. Alternative heating mechanisms include, for example, resonant Landau damping of kinetic Alfv\'en waves \citep{2018JPlPh..84a9005H,2019NatCo..10..740C,2022ApJ...924L..26C}, and stochastic perpendicular heating of ions resulting from the violation of the magnetic moment conservation induced by large-amplitude turbulent fluctuations at ion scales \citep{2010ApJ...720..503C,2020ApJS..246...30M}. These mechanisms, however, are not discussed here.%Finally, the observation of high perpendicular temperatures in both the corona and the solar wind is also more consistent with the ion cyclotron picture.

In the attempt to reconcile the remote-sensing and in-situ observations in a unified view of the heating of the coronal plasma and the subsequent acceleration of the solar wind, local measurements of the coronal regions are needed. At present, PSP has only occasionally entered the solar corona \citep{2021PhRvL.127y5101K,2022ApJ...926L...1B,2022ApJ...926L..16Z,2022ApJ...934L..36Z,marino20231}, although its progressively shrinking orbits will ensure that the spacecraft spends extended periods immersed in the coronal plasma. Even then, though, single-spacecraft measurements will provide only one-point information, precluding a clear understanding of how the global (magnetic) configuration of coronal sources affects local dynamics and turbulence properties in space plasmas. It appears thus that remote-sensing observations of large-scale source regions of the solar wind are of critical importance to conduct this type of investigations. Correspondingly, these alone offer only some general, though not conclusive, insights. It follows that exploiting the synergies between in-situ measurements and remote-sensing observations and, in particular, the joint PSP -- SO observations of the Sun and its environment, is the most viable approach to answer key questions concerning the coronal heating. Two companion papers by \citet{2021ApJ...920L..14T,2022ApJ...935..112T} can be considered a first attempt to exploit observations of Metis \citep[the coronagraph aboard SO;][]{2020ExA....49..239F,2020A&A...642A..10A} in combination with in-situ PSP measurements collected while in orbital quadrature, to link global and local properties of coronal flows and study their evolution in the transition from sub- to super-Alfv\'enic wind.

The first Metis observations, performed in $2020$, imaged the solar corona in its quasi-dipolar magnetic configuration typical of the minimum activity phase out to $\sim7$ R$_{\odot}$, thus allowing the extension of the study of the morphology and dynamics of the coronal Current Sheet (CS), that originates at the cusp of the equatorial streamers, beyond the region explored with UVCS and limited to a heliocentric distance of $5$ R$_{\odot}$ \citep[see the review by][]{2020SSRv..216..117A}. In the Metis observations, the initial coronal part of the heliospheric current sheet, separating opposite polarities of the coronal magnetic field, is clearly outlined by a slightly warped quasi-equatorial layer, of denser and slower wind plasma that is undergoing a significant acceleration between $4$ and $5$ R$_{\odot}$ \citep{POP22-AR-PPSEP2023-01576}. These first Metis observations show that the zone where the slow wind is observed in corona, $\lesssim30^{\circ}$ in latitude, is structured as follows: the slowest wind, observed in a layer $<10^{\circ}$ wide in latitude and associated with the current sheet, is surrounded by steep velocity gradients, a few degrees wide in latitude, indicating a transition to slow wind flows characterized by higher velocity but lower acceleration relative to plasma directly embedded in the current sheet \citep{POP22-AR-PPSEP2023-01576}.

This paper presents evidence on the occurrence of ICWs in the coronal structures explored by PSP, taking advantage of the SO/Metis -- PSP joint observations of mid-January $2021$. Specifically, the results show that when PSP flies through the layers of the heliospheric CS remotely imaged by Metis in its initial coronal stretch, turbulence increases significantly causing the plasma to depart more from an equilibrium state. A case of local PSP measurements along the CS, in which a proton cyclotron instability driven by temperature anisotropy generates ICWs, is shown in the present work. This combined observation, the first of its kind to be reported, represents a step forward in understanding how coronal large-scale structures drive and regulate small-scale fluctuations/waves in the space plasma. This paper first presents a description of the coordinated SO/Metis -- PSP observations (\S{} \ref{sec:observations}), then the methodological approach implemented to link local measurements of the solar wind to remotely imaged coronal source regions (\S{} \ref{sec:analysis}), the observation of the onset of ICWs along the turbulent CS (\S{} \ref{sec:results}), and finally concluding remarks on the results presented (\S{} \ref{sec:conclusions}).

%Evidence that Alfv\'en cyclotron waves are present only at some specific coronal regions places strong constraints on models that aim to explain the heating/acceleration of the slow solar wind via either resonant or non-resonant interaction with plasma particles.%

\section{SO/Metis -- PSP joint observations}
\label{sec:observations}
\begin{figure*}
	\begin{center}
		\includegraphics[width=\linewidth]{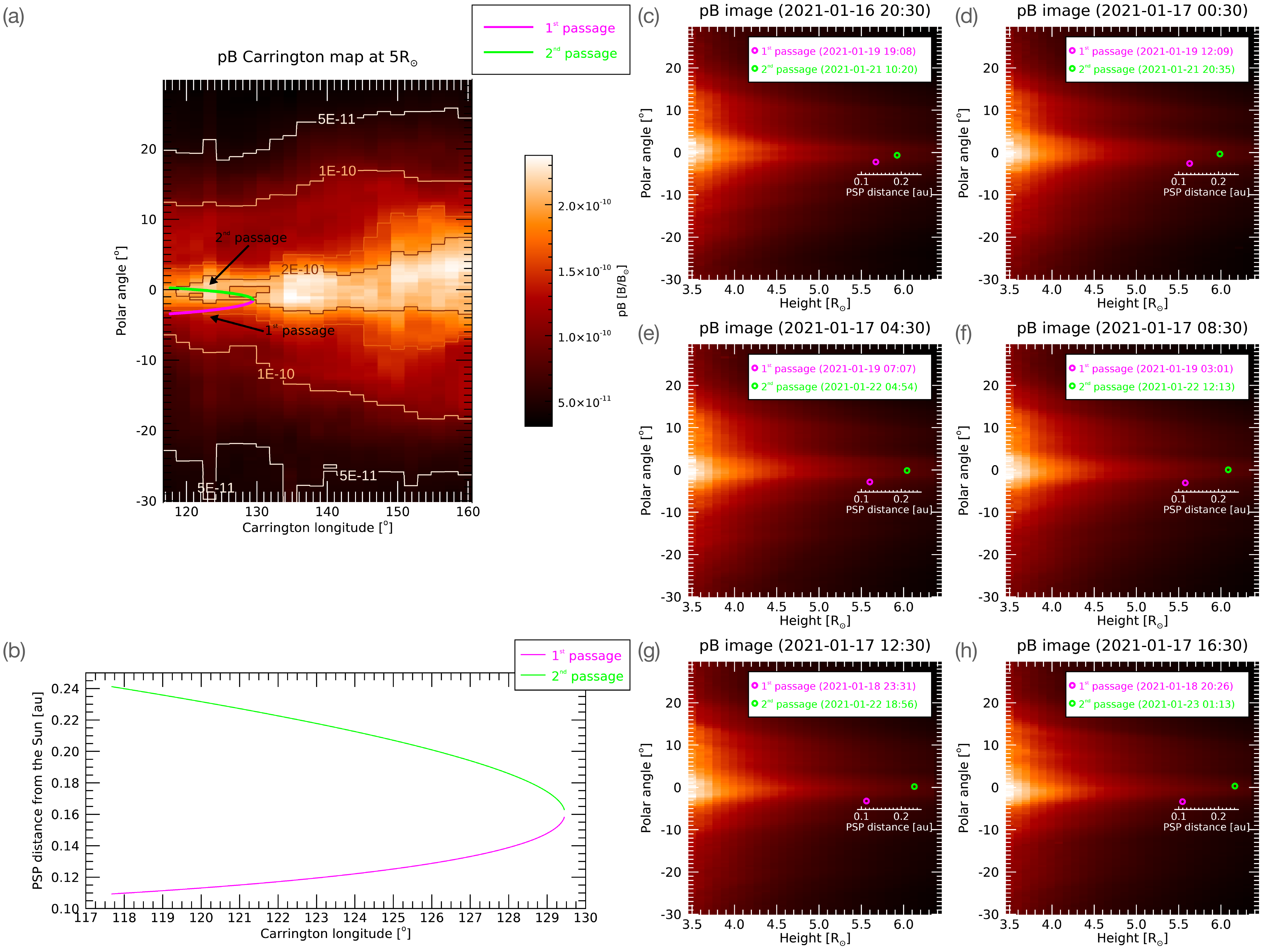}
	\end{center}
	\caption{Overview of the joint SO/Metis -- PSP observations. (a) Carrington map at $5$ R$_{\odot}$ from Metis pB observations for the eastern solar limb on which is superimposed the back-projected trajectory of PSP during the first (magenta) and second (lime) passage (also labeled). (b) PSP distance as a function of Carrington longitude in the first (magenta) and second (lime) crossing of the same POS observed by Metis. (c)--(h) Zoom of the FOV observed by Metis above the East limb of the Sun in the last $6$ images acquired by the coronagraph; the apparent position of PSP during the Metis POS crossing is marked by an open circle (the color code is the same as the previous panels); the times at which the crossings occur are also reported in the legend; the extra white x-axis indicates the actual distance of PSP from the Sun.}
	\label{fig:metis_carrington_map}
\end{figure*}

From January $14$ through January $17$, $2021$, Metis acquired, with $4$-hour cadence, $23$ images of the white-light solar corona in an annular Field Of View (FOV) from $3.5$ to $6.3$ R$_{\odot}$, which were calibrated in flight according to \citet{ref:calibration_vl}. Figure \ref{fig:metis_carrington_map}(a) shows the Carrington map (i.e., a synoptic chart displayed using Carrington coordinates) of the polarized Brightness (pB) at a distance of $5$ R$_{\odot}$ above the East limb of the Sun.

The coronal CS, namely the initial part of the heliospheric current sheet that is imaged at coronal level, is clearly visible in the Metis-based Carrington map around the solar equator as the densest structure. In particular, the dark orange level line at $2\times10^{-10}$ B$/$B$_{\odot}$ (which helps to roughly outline the CS boundaries) shows the apparent presence of a double structure for longitudes less than $\sim132^{\circ}$. As discussed in \citet{2021ApJ...920L..14T}, this is due to two small CS warps extending up to $\sim\pm10^{\circ}$ in latitude. Due to its retrograde motion relative to the Sun's rotation, PSP flies through the Plane Of the Sky (POS) observed by Metis twice \citep[and later to the Metis observation, see][for a more extensive discussion on the PSP orbital characteristics]{2021A&A...650A..21S,2022ApJ...935..112T}. The PSP back-projected trajectory on the pB Carrington map is shown in Figure \ref{fig:metis_carrington_map}(a) with magenta and lime lines, respectively, for the first and second crossing. PSP collected measurements around the CS, while moving away from the Sun from $0.11$ to $0.24$ au (Figure \ref{fig:metis_carrington_map}(b)). The apparent location of PSP with respect to the CS observed by Metis is shown in Figures \ref{fig:metis_carrington_map}(c)--(h): during its orbit, PSP moved in latitude and sampled different portions of the CS.

\section{Linking PSP measurements to Metis observations of coronal sources}
\label{sec:analysis}
\begin{figure*}
	\begin{center}
		\includegraphics[width=\linewidth]{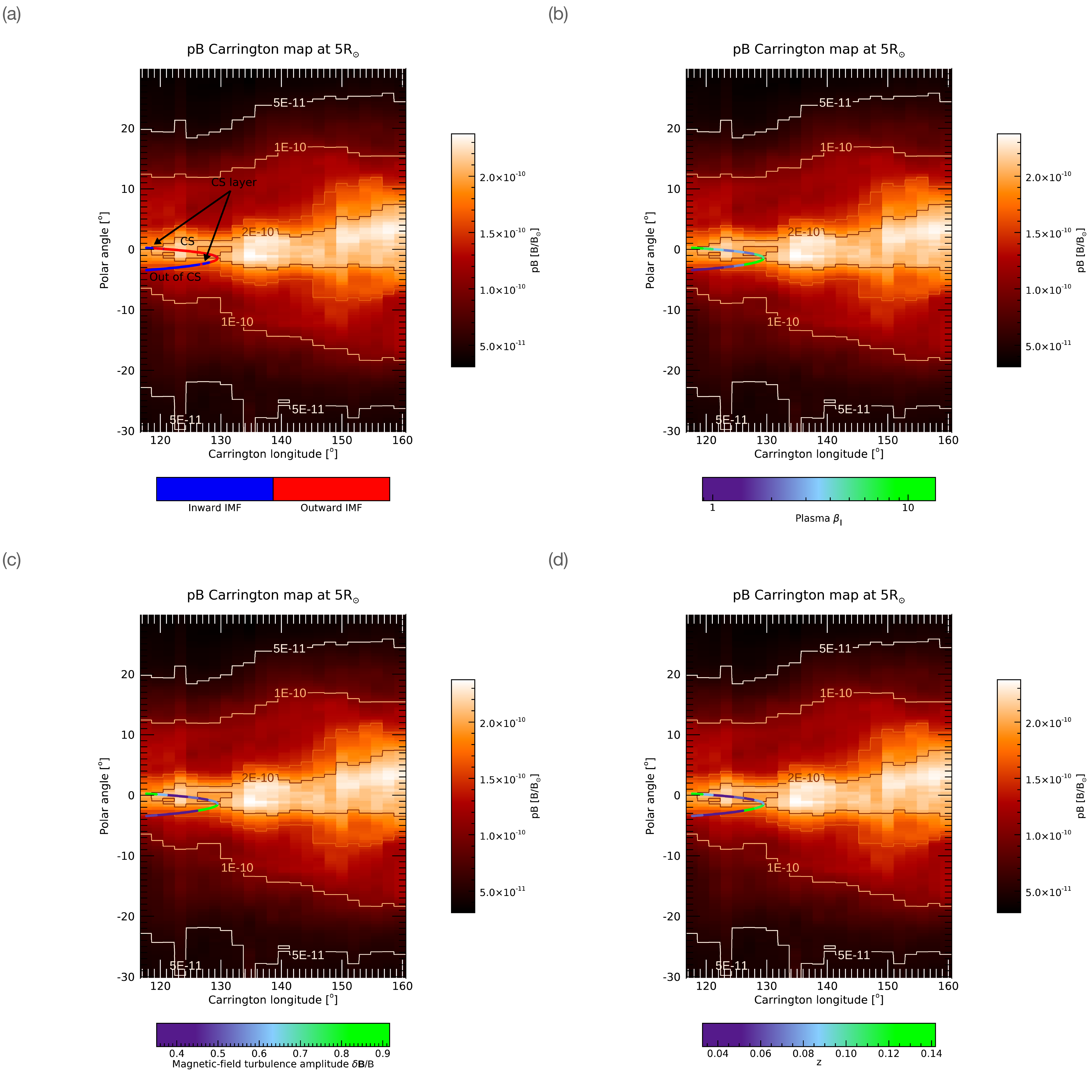}
	\end{center}
	\caption{Projection onto the Metis pB Carrington map of the IMF direction (a), proton plasma $\beta_{\parallel}$ (b), magnetic-field turbulence amplitude $\delta\mathbf{B}/\langle B\rangle$ (c), and magnetic-field deflection $z$ (d), estimated from PSP measurements. The topology of the CS as flown through by PSP is outlined in (a).}
	\label{fig:psp_on_metis_observations}
\end{figure*}

Local solar wind measurements of PSP can thereby be related to the coronal source structures imaged by Metis. Figure \ref{fig:psp_on_metis_observations} shows some relevant PSP-based MHD quantities projected onto the Metis pB Carrington map in $1.68^{\circ}$-wide bins. Specifically, panels (a)--(d) display the direction of the Interplanetary Magnetic Field (IMF), the proton plasma $\beta_{\parallel}=\upsilon^{2}_{\mathrm{th}_{\parallel}}/\upsilon^{2}_{\mathrm{A}}$ (where $\upsilon_{\mathrm{th}_{\parallel}}$ and $\upsilon_{\mathrm{A}}$ are the parallel proton thermal and Alfv\'en speeds, respectively), the magnetic-field turbulence amplitude $\delta\mathbf{B}/\langle B\rangle=\sqrt{\langle|\mathbf{B}(t)-\langle\mathbf{B}\rangle|^{2}\rangle}/\langle B\rangle$ (where $\langle...\rangle$ indicates time averaging over the typical fluid scale of $10$ minutes\footnote{Note that the results are the same at other scales in the inertial range.}), and the magnetic-field deflection $z=(1-\cos\alpha)/2$ \citep[where $\alpha$ is the angle between the local and mean magnetic fields,][]{2020ApJS..246...39D}. Magnetic field and plasma data come from the fluxgate magnetometer and the top-hat electrostatic analyzer of the FIELDS \citep{2016SSRv..204...49B} and Solar Wind Electrons Alphas \& Protons \citep[SWEAP;][]{2016SSRv..204..131K} instrument suites on board PSP, respectively.

PSP measurements of the IMF direction (Figure \ref{fig:psp_on_metis_observations}(a)) agree with the topological structure of the CS imaged by Metis. Specifically, the polarity reversal (which identifies the CS boundaries) occurs in the proximity of the dark orange level line that helps to outline the CS edges. Thus, it can be deduced from PSP observations that the spacecraft, initially outside of the CS (in blue), first entered the CS at Carrington longitude of about $126^{\circ}$ (in red), and crossed the CS a second time, exiting it at Carrington longitude of about $119^{\circ}$. A schematic of the magnetic topology of the CS thus inferred is illustrated in Figure \ref{fig:psp_on_metis_observations}(a). At the double crossing of the CS (occurring at the ``CS layers'', Figure \ref{fig:psp_on_metis_observations}(a)), the plasma $\beta_{\parallel}$ increases by an order of magnitude compared with the region outside (Figure \ref{fig:psp_on_metis_observations}(b)). This is expected when considering that the current sheet layers (dominated by opposite parallel magnetic currents) are sites of magnetic reconnection events, likely related, e.g., to the tearing mode instability \citep{1985RPPh...48..955P,1994SSRv...70..299V,2022GeoRL..4996986P}, which tends to form magnetic islands, i.e., small flux ropes, disconnecting from the CS \citep{2021A&A...650A..12Z,2022A&A...659A.110R}. At the CS boundaries, the magnetic-field turbulence also increases (Figure \ref{fig:psp_on_metis_observations}(c)), mostly due to the change in field direction, as indicated by the concurrent increase in the parameter $z$ (Figure \ref{fig:psp_on_metis_observations}(d)). Recent analysis of PSP measurements shows that turbulence is typically reduced in the proximity of the heliospheric CS \citep[see, e.g.,][and references therein]{2021A&A...650L...3C}. In particular, observations suggest that closer to the CS (i.e., for angular distance from the CS within $\simeq4^{\circ}$) fluctuations are similar to non-Alfv\'enic, slow solar wind (with steeper and broader power-law spectra, smaller fluctuation amplitude), while they are closer to Alfv\'enic, fast wind further away. \citet{2020ApJS..246...26Z,2021A&A...650A..12Z} presented a detailed analysis of PSP data in the vicinity of the CS, identifying many magnetic islands, which they argue are consistent with non-Alfv\'enic turbulence. These are preferentially located in the vicinity of the CS. \citet{2022ApJ...928...93S} also report, exploiting MHD simulations complemented by analysis of in-situ data at $1$ au, the decrease in Alfv\'enicity of turbulence around the heliospheric CS. The present observations lie within the angular range considered as CS in \citet{2021A&A...650L...3C}, and therefore show finer details of the turbulence structure in its boundaries. This result is consistent with the following scenario. At the CS, the solar wind is slower than typical of other coronal regions. Nevertheless, the regions along it (i.e., its edges) are characterized by faster (though still slow) coronal flows and lower wind acceleration. This is clearly shown by \citet{POP22-AR-PPSEP2023-01576}. It follows that the CS layers might be sites of velocity shears. These are also implied by the magnetic field configuration, the velocity being inversely correlated with the expansion factor, which is maximum at the CS. Velocity shears are known to generate turbulence \citep{2016JGRA..12111021S,2019PhRvL.122c5102S} and deflect the magnetic field \citep[possibly in association with switchback formation,][]{2020ApJ...902...94R} through the Kelvin-Helmholtz (KH) instability. Even without invoking KH-driven magnetic fluctuations, the small-scale flux ropes disconnected from the CS by magnetic reconnection enhance the amplitude of the quasi-$2$D MHD turbulence \citep{2018ApJ...854...32Z,2020ApJ...900..115Z}.

\begin{figure*}
	\begin{center}
		\includegraphics[width=\linewidth]{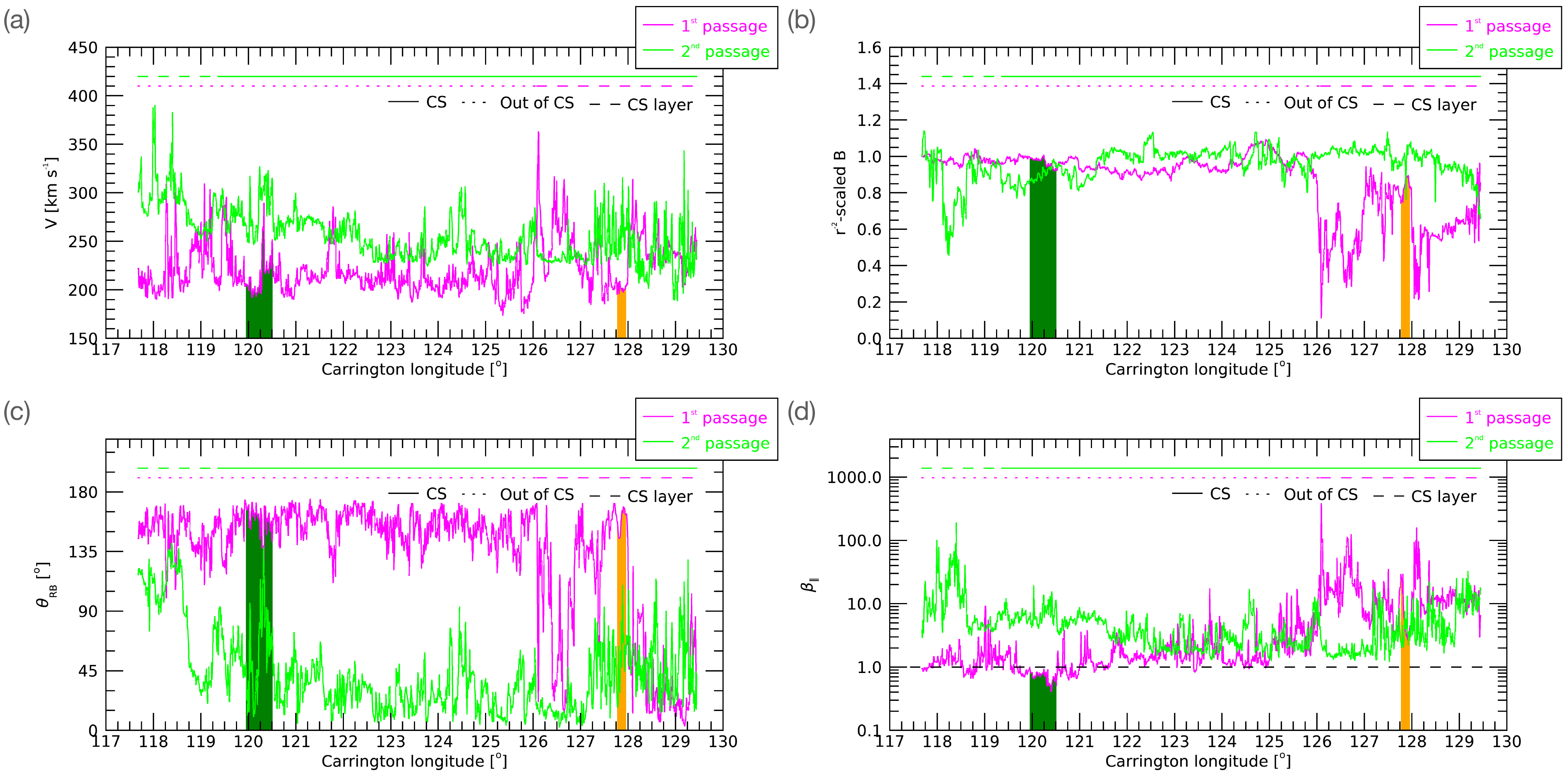}
	\end{center}
	\caption{Measurements of flow speed $V$ (a), $r^{2}$-scaled magnetic field intensity (b), magnetic field-radial direction angle $\theta_{RB}$ (c), and proton plasma $\beta_{\parallel}$ (d) against Carrington longitude, during the first (magenta) and second (lime) PSP passage through the POS observed by Metis. Horizontal lines with different styles schematically depict the topology of the CS crossed by PSP. The green (orange) shaded area denotes an interval outside (at the edge) of the CS analyzed in \S{} \ref{sec:results}.}
	\label{fig:longitude_series}
\end{figure*}

In order to clarify the correlation between the PSP measurements and the different regions of the CS crossed by the spacecraft, to provide additional information, such as the extent of the CS layers, and to support/complement the results outlined by Figure \ref{fig:psp_on_metis_observations}, Figures \ref{fig:longitude_series}(a)--(d) show profiles of solar wind speed $V$, magnetic field magnitude $B$ normalized to the square of the distance $r$ from the Sun\footnote{Since $\nabla\cdot\mathbf{B}=0$ the radial component of the magnetic field, $B_{R}$, clearly scales as $r^{2}$, and as long as $B_{R}$ is the dominant contributor to $\mathbf{B}$-magnitude as during these observations, so does $B$.}, angle between magnetic field vector and radial direction $\theta_{RB}$, and proton plasma $\beta_{\parallel}$, as a function of the Carrington longitude range observed in conjunction by PSP and Metis.

As mentioned above, the boundaries of the CS were identified as the regions around the magnetic field reversal. More specifically, Figure \ref{fig:longitude_series}(c) clearly shows that during the first PSP passage, the magnetic field is directed towards the Sun out to $\sim126^{\circ}$ Carrington longitude ($\theta_{RB}\sim180^{\circ}$). Therefrom, the magnetic field shows no definite direction, fluctuating from inward ($\theta_{RB}\sim180^{\circ}$) to outward ($\theta_{RB}\sim0^{\circ}$) orientation. This is the region of the CS layer. Similarly, during the second PSP passage, from an initial positive polarity ($\theta_{RB}\sim0^{\circ}$), the magnetic field changes direction for Carrington longitudes less than $\sim119^{\circ}$, thereby identifying the second CS layer. It appears evident that at the edges of the CS (marked by dashed horizontal lines in the panels of Figure \ref{fig:longitude_series}), the plasma speed, while remaining slow, undergoes moderate acceleration (Figure \ref{fig:longitude_series}(a)), albeit more pronounced during the second crossing (lime line). This might be indicative of the presence of velocity shears that could explain the onset of greater turbulence at the CS boundaries. The latter is evident from the larger amplitude of $B$ fluctuations (Figure \ref{fig:longitude_series}(b)) as well as the more intense and numerous magnetic field deflections, shown by the larger fluctuations of the $\mathbf{B}$ inclination with respect to the sampling direction. Finally, based on the aforementioned proposed mechanisms, the plasma $\beta_{\parallel}$ also increases significantly at the CS layers (Figure \ref{fig:longitude_series}(d)). In addition to helping identify the different portions of the CS, the longitudinal series also allow for a more accurate estimate of the longitudinal extent of its boundaries, which turns out to be $3.35^{\circ}$ and $1.67^{\circ}$ wide during the first and second crossing, respectively.

\section{Onset of ICWs along the turbulent CS}
\label{sec:results}
\begin{figure*}
	\begin{center}
		\includegraphics[width=\linewidth]{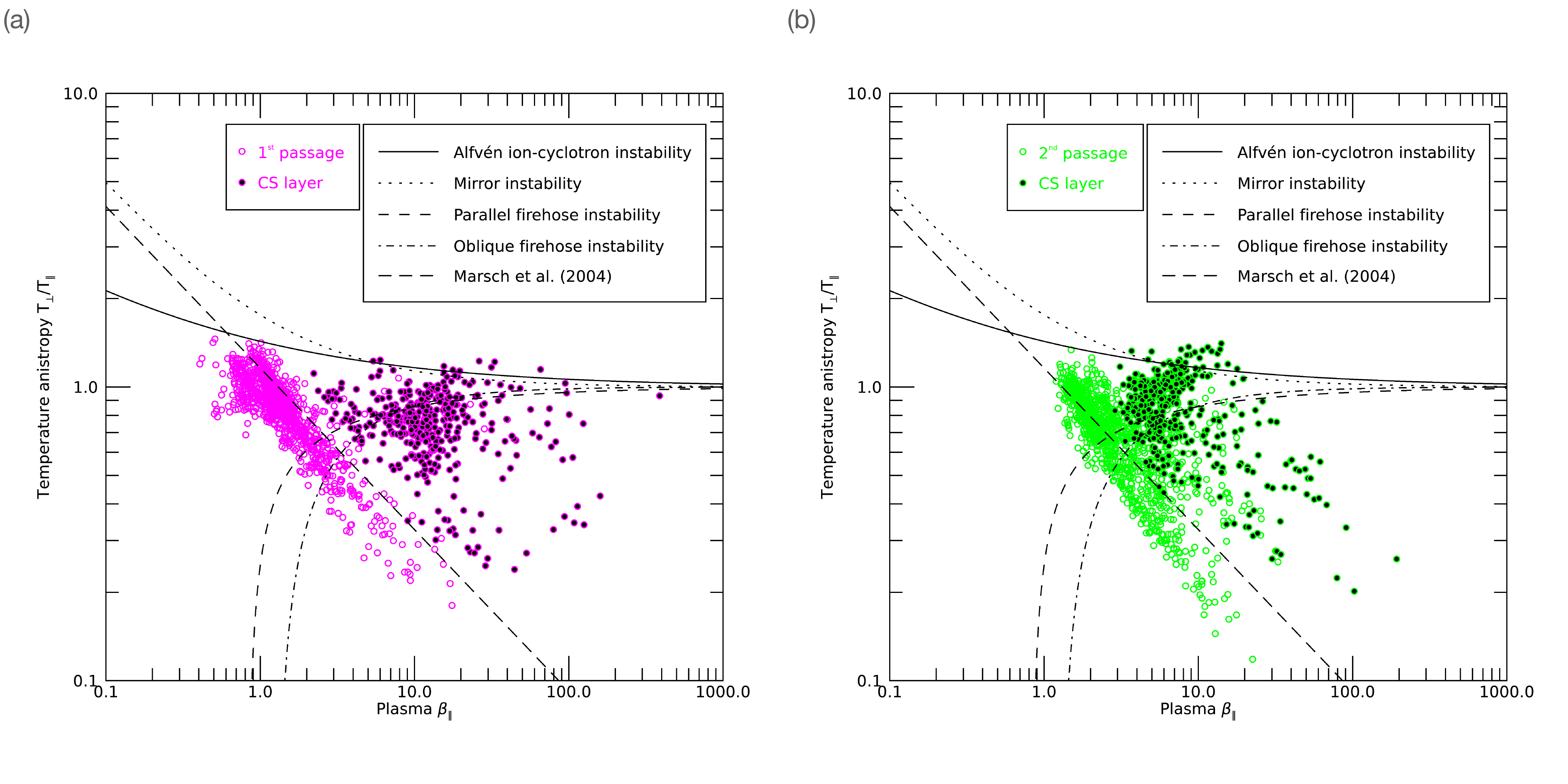}
	\end{center}
	\caption{Scatterplot of the PSP observations in the $\beta_{\parallel}-T_{\perp}/T_{\parallel}$ plane during the first (a) and second (b) PSP passage through the POS observed by Metis. The Alfv\'en ion-cyclotron, mirror-mode, parallel and oblique firehose instability threshold curves are displayed as solid, dotted, dashed, and dash-dotted lines, respectively. The empirical inverse relation by \citet{2004JGRA..109.4102M} is also shown as a long-dashed line. Black filled symbols refer to regions at the edges of the CS.}
	\label{fig:brazil_plot}
\end{figure*}

In order to assess whether the increase in turbulence occurring in the plasma across the CS affects the kinetic properties of the particle velocity distribution functions, especially regarding the nature of unstable wave-particle interactions, PSP measurements are plotted in the $\beta_{\parallel}-T_{\perp}/T_{\parallel}$ plane (where $T_{\perp}/T_{\parallel}$ is the temperature anisotropy) in Figure \ref{fig:brazil_plot}. The distribution is compared with different plasma instability thresholds (marked with different line types) as estimated by \citet{2006GeoRL..33.9101H} for a maximum growth rate $\gamma=10^{-3}\Omega_{p}$, where $\Omega_{p}$ is the proton gyro-frequency \citep[see also the comprehensive review by][and references therein]{2019LRSP...16....5V}.

When separating the data points based on the different portions of the CS crossed by PSP as identified in Figures \ref{fig:psp_on_metis_observations} and \ref{fig:longitude_series}, two different populations are distinguishable (though, less evident during the second passage). When inside or outside the CS (open symbols) PSP data distribute fairly well along the empirical anti-correlation curve found by \citet{2004JGRA..109.4102M} with Helios measurements (long-dashed line in Figure \ref{fig:brazil_plot}). Observations corresponding to the CS layers (black filled symbols) are instead mainly located at larger plasma $\beta_{\parallel}$. All instability thresholds shown in Figure \ref{fig:brazil_plot} decrease with increasing $\beta_{\parallel}$. Therefore, the distribution of the CS layer data lies closer to the unstable regions of parameter space. Indeed, a significant portion of the parameter values computed from the data collected during the second CS crossing (Figure \ref{fig:brazil_plot}(b)) extends beyond the thresholds. As a result, the plasma along the CS is more likely to be disequilibrated and thus trigger wave-particle instabilities. Therefore, it can be argued that the increased turbulence experienced at the CS boundaries, due to either magnetic reconnection-driven $2$D flux ropes \citep{2020ApJS..246...26Z,2021A&A...650A..12Z} or KH-fluctuations \citep{2022ApJ...929...98T}, indirectly triggers the Alfv\'en ion-cyclotron and the firehose instability via changes of the particle distributions. These instabilities then transfer energy from the particles (and thus restore a Maxwellian equilibrium condition) to the electromagnetic fields, i.e., generating ICWs and fast-mode waves \citep[e.g.,][]{2019PhRvL.122c5102S}.

\begin{figure*}[h]
	\begin{center}
		\includegraphics[width=\linewidth]{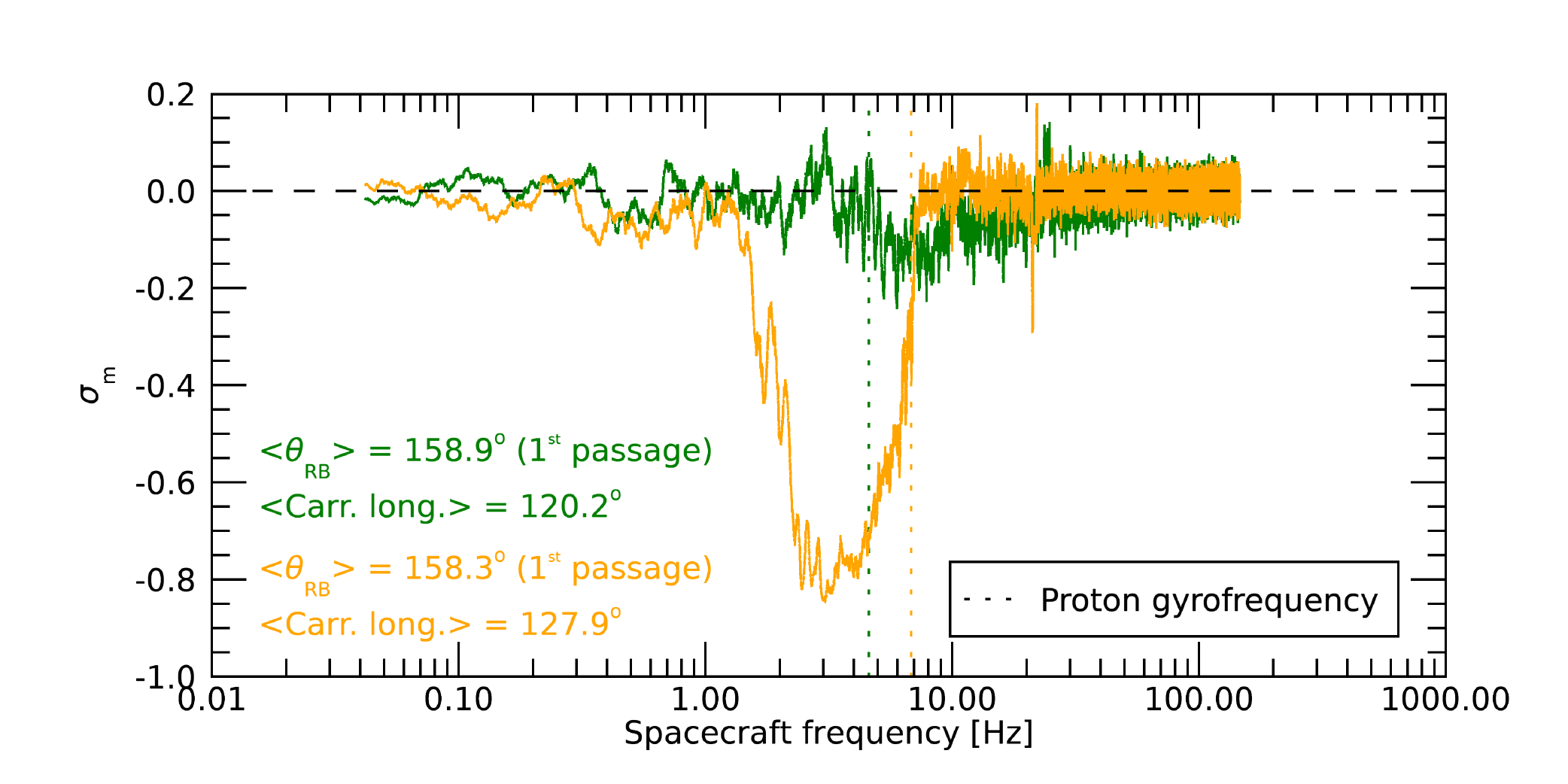}
	\end{center}
	\caption{Magnetic helicity $\sigma_{m}$ spectra computed outside (green) and along (orange) the CS when PSP was sampling the nearly parallel solar-wind magnetic field marked by same-color shaded regions in Figure \ref{fig:longitude_series}. The corresponding local proton gyro-frequencies $\Omega_{p}$ are indicated by color-coded vertical dotted lines. The average Carrington longitude and sampling angle during both observations are also reported.}
	\label{fig:icws}
\end{figure*}

The presence of modes driven by temperature-anisotropy instabilities can in principle be inferred by identifying in space plasmas the corresponding polarization properties at the characteristic frequencies. Specifically, the presence of left-handed circularly polarized ICWs, which result from the Alfv\'en ion-cyclotron instability, can be indicated through a strong peak in the normalized magnetic helicity $\sigma_{m}$ spectrum near the local proton gyro-frequency \citep[the magnetic helicity can be indeed be used to diagnose the wave polarization state,][]{1982JGR....8710347M,1982PhRvL..48.1256M}. Moreover, the deeper and wider the helicity peak, the more numerous are the waves \citep{2019ApJ...885L...5T}. Searching for ICWs in the solar wind is, however, tricky. These are indeed waves with a wavevector parallel to the magnetic field. This means that any spacecraft can detect this kind of wave only when sampling solar wind plasma parallel to the mean magnetic field. One-hour intervals were then selected, inside, outside and along the CS, where the $\sigma_{m}$ spectrum was calculated, looking for significant peaks around $\Omega_{p}$. It transpires that on average ICWs are prevalent at the edges of the CS, but are relatively scarce (about $50$\% less abundant) inside or outside the CS. Examples are given in Figure \ref{fig:icws}, where the $\sigma_{m}$ spectra obtained in one-hour intervals, nearly parallel to the magnetic field, outside the CS (green) and along its boundaries (orange), during the first PSP trajectory, are shown. These intervals are indicated by shaded areas (with the same color code) in the longitude series in Figure \ref{fig:longitude_series}. Note that because the spacecraft orbital velocity is greater during the first passage (PSP is closer to the Sun) the range of longitudes spanned in an hour is larger.

Although it should be mentioned that a higher Alfv\'enic contribution at fluid scales can also lead to the onset of ICWs \citep[the reader is referred, in this regard, to][]{2015ApJ...811L..17B,2019ApJ...885L...5T} even far from the CS, the analysis reported here suggests that instabilities, including the ICWs as shown in the example of Figure \ref{fig:icws}, are driven along the CS, where turbulent phenomena, related to either KH instability or magnetic disconnection processes, are most intense.

A large fraction of the data points lie above the thresholds for the firehose instabilities (dashed and dash-dotted lines, respectively, in Figure \ref{fig:brazil_plot}). Although this applies both inside and outside the CS, it occurs mainly along the CS boundaries (only the core population at the CS layers exceeds the firehose instability; this is more evident during the first passage). In these time intervals, the system can drive parallel fast-mode waves with a polarization opposite from ICWs in the plasma frame \citep{2021ApJ...912..101W}. A further in-situ analysis of these fast-mode waves, however, is beyond the scope of the present work though, which focuses primarily on ICWs, whose presence is evidenced by Figure \ref{fig:icws} and whose importance is related to their possible contribution in the coronal plasma heating mechanisms discussed in \S{} \ref{sec:introduction}.

\section{Concluding remarks}
\label{sec:conclusions}
The linkage of the temporal pattern of the increased turbulence and ICW burst with the spatial topology of the coronal structures documented in Figures \ref{fig:metis_carrington_map}--\ref{fig:icws} suggests this possible scenario. The large-scale CS layers are source regions of turbulence (via either KH instability or magnetic reconnection-driven $2$D flux ropes), which pushes the plasma above the thresholds of instabilities, which in turn drive small-scale fluctuations/waves and regulate the plasma. Obviously, these effects are very strong near the Sun, where this is therefore a very efficient re-shuffling of energy. Moreover, the correlation between the increased turbulence and ion cyclotron modes is important for connecting turbulence directly to heating. By estimating the turbulence energy per volume and adding up the energy in the ion cyclotron modes per unit volume and further assuming this is dissipated (at a later time) over an ion inertial time scale, it would indeed be possibile to obtain the heating rate and estimate the expected temperature. Theoretical work is currently in progress to test whether the energy released by turbulence and ICWs is sufficient to heat the plasma to temperatures of a million degrees and accelerate the slow solar wind along the CS. In addition, it would be worthwhile to undertake a more comprehensive statistical study of the occurrence of small-scale fluctuations inside and outside observed CS layers to obtain a better understanding of the universality of the proposed scenario. It is finally worth noting that the present results show a link between increased turbulence along the CS layers and the onset of temperature anisotropy-driven instabilities, without however proving (but only suggesting) a cause-and-effect relationship. Causality between increased levels of turbulence and an increased strength of non-equilibrium plasma features could only be evidenced through statistical work and a non-linear treatment of the turbulence-plasma interactions, which is left for future work.

This result shows the tremendous science potential of the Metis coronagraph on board SO, especially in conjunction with PSP. Indeed, the need for comparing/combining remote observations of the extended corona with local measurements of the solar wind acquired by different spacecraft (in quadrature) emerges clearly, not only to understand how large-scale coronal dynamics drives the onset of turbulence, waves, and kinetic instabilities, but also to develop simulation-based numerical experiments and/or theoretical models that can tackle long-lasting questions such as the coronal heating and wind acceleration problem.

\acknowledgments
Solar Orbiter is a space mission of international collaboration between ESA and NASA, operated by ESA. D.T. was partially supported by the Italian Space Agency (ASI) under contract 2018-30-HH.0. G.P.Z., L.A., and L.-L.Z. acknowledge the partial support of a NASA Parker Solar Probe contract SV4-84017, an NSF EPSCoR RII-Track-1 Cooperative Agreement OIA-2148653, and a NASA IMAP grant through SUB000313/80GSFC19C0027. L.S.-V. was funded by the SNSA grants 86/20 and 145/18. D.V. is supported by STFC Ernest Rutherford Fellowship ST/P003826/1 and STFC Consolidated Grants ST/S000240/1 and ST/W001004/1. The Metis program is supported by ASI under contracts to the National Institute for Astrophysics and industrial partners. Metis was built with hardware contributions from Germany (Bundesministerium f\"ur Wirtschaft und Energie through the Deutsches Zentrum f\"ur Luft- und Raumfahrt e.V.), the Czech Republic (PRODEX) and ESA. The Metis data analyzed in this paper are available from the PI on request. Parker Solar Probe data was downloaded from the NASA's Space Physics Data Facility (\href{https://spdf.gsfc.nasa.gov}{https://spdf.gsfc.nasa.gov}). This work was discussed at the ISSI Team ``Ion Kinetic Instabilities in the Solar Wind in Light of Parker Solar Probe and Solar Orbiter Observations'' led by L. Ofman and L. Jian.
\par

%\bibliography{reference_list}{}
%\bibliographystyle{aasjournal}

\end{document}